\documentclass[prd,preprintnumbers,superscriptaddress,nofootinbib,twocolumn]{revtex4-2}
\usepackage{epsfig}
\usepackage{amsmath}
\usepackage{amsfonts}
\usepackage{float}
\usepackage{amssymb}
\usepackage{slashed}
\usepackage{color}
\usepackage{pbox}
\usepackage{subfigure}
\usepackage[colorlinks,citecolor=blue, linkcolor = blue, urlcolor = blue]{hyperref}
\usepackage{tabularx}
\usepackage{titlesec}
\usepackage{scrextend}
\usepackage[normalem]{ulem}
\usepackage{orcidlink}
\usepackage{comment}
\usepackage{MnSymbol}

\begin{document}


\title{Inelastic Singlet-Doublet Fermion Dark Matter in light of the 248 keV LZ event}

\author{{Debasish Borah\orcidlink{https://orcid.org/0000-0001-8375-282X}}}
\email{dborah@iitg.ac.in}
\affiliation{Department of Physics, Indian Institute of Technology Guwahati, Assam 781039, India}
\author{Sujit Kumar Sahoo{\orcidlink{https://orcid.org/0000-0002-9014-933X}}}
\email{ph21resch11008@iith.ac.in}
\affiliation{Department of Physics, Indian Institute of Technology Hyderabad, Kandi, Sangareddy 502285, Telangana, India}
\affiliation{Institute of Mathematical Sciences (IMSc), Chennai 600113, India}
\author{Narendra Sahu{\orcidlink{https://orcid.org/0000-0002-9675-0484}}}
\email{nsahu@phy.iith.ac.in}
\affiliation{Department of Physics, Indian Institute of Technology Hyderabad, Kandi, Sangareddy 502285, Telangana, India}
\author{Shashwat Sharma{\orcidlink{https://orcid.org/0009-0002-2266-0467}}}
\email{ph23resch11016@iith.ac.in}
\affiliation{Department of Physics, Indian Institute of Technology Hyderabad, Kandi, Sangareddy 502285, Telangana, India}

\begin{abstract}
Recently, the LUX-ZEPLIN (LZ) collaboration reported the observation of a single dark matter (DM)-nucleus scattering event at a nuclear recoil energy of $248\pm23_{\rm stat}\pm23_{\rm sys}$ keV, corresponding to an exposure of 2.84 tonne-year. The absence of events at lower nuclear recoil energies in the predicted spectrum is naturally explained if the underlying process is inelastic DM-nucleus scattering. Motivated by this, we investigate the singlet-doublet fermion DM model, in which the DM consists of two pseudo-Dirac states: the Majorana nature of the lighter state forbids tree level $Z$-mediated elastic scattering identically, while the same states enable inelastic DM-nucleus scattering via $Z$ exchange, with any residual elastic scattering proceeding only through a suppressed Higgs-mediated channel. We further extend the model with a $Z_2$-even scalar triplet, which is responsible both for generating the pseudo-Dirac splitting and for realizing Majorana neutrino masses via the Type-II seesaw
mechanism.
\end{abstract}
\maketitle

\section{Introduction} 
\label{sec:intro}
Recently, the LUX-ZEPLIN (LZ) collaboration has reported a single high-energy nuclear recoil event using 2.84 tonne-yr of exposure, corresponding to nuclear recoil energy of $248 \pm 23 (\rm stat) \pm 23 (\rm sys)$ keV. This event is in tension with the background-only hypothesis at a global significance of $2.6\sigma$ \cite{LZ:2026axp} and can reach a maximal local significance of $3.4\sigma$. The event can be described by weakly interacting massive particle (WIMP) dark matter (DM) with masses above 200 GeV undergoing inelastic up-scattering off nucleons. While the statistical significance is far from the discovery limit, this is still tantalizing due to its appearance in a low-background regime and has already motivated several particle physics interpretations like Higgsino DM \cite{Fan:2026kxx, Wu:2026nhi, Freese:2026sga, Du:2026guj, Pospelov:2026ewn, Rodd:2026tyn}, other DM candidates \cite{McCabe:2026crm, Unwin:2026rdp, Smirnov:2026aqk, Nomura:2026qyq, Lou:2026idn, Su:2026rwz, DiMauro:2026ldr, Yamashita:2026ump, Chattopadhyay:2026ryw, deLima:2026shq, Visinelli:2026kgt}, DM, in general \cite{Dent:2026bji, Gu:2026vto} or even non-DM origin \cite{Jeesun:2026vzo}. 

Motivated by this, we study one of the popular WIMP DM scenario known as the singlet-doublet DM (SDDM)\cite{Bhattacharya:2018fus,Cynolter:2015sua,Bhattacharya:2015qpa,Bhattacharya:2017sml,Bhattacharya:2018cgx,Bhattacharya:2016rqj,Dutta:2020xwn,Borah:2021khc,Borah:2021rbx,Borah:2022zim,Borah:2023dhk,Paul:2024iie,Paul:2026snc,DEramo:2007anh,Cohen:2011ec,Freitas:2015hsa,Calibbi:2015nha,Cheung:2013dua,Banerjee:2016hsk,DuttaBanik:2018emv,Horiuchi:2016tqw,Restrepo:2015ura,Abe:2017glm,Konar:2020wvl,Konar:2020vuu,Calibbi:2018fqf,Ghosh:2021wrk,Das:2023owa,Bhattacharya:2021ltd,Enberg:2007rp,Oncala:2021tkz,Paul:2024prs,Paul:2025spm,Dey:2025pcs,Borah:2026hpp,Restrepo:2019soi,Restrepo:2022cpq,Bhattiprolu:2026hnp,Bhattiprolu:2025beq} in the context of the 248 keV recoil event reported by the LZ collaboration. While DM in the minimal vector-like singlet-doublet fermion setup has diagonal interactions mediated by $Z$ boson, it is possible to make the $Z$-mediated interactions off-diagonal by introducing Majorana mass terms. As the required mass splitting is small $\sim \mathcal{O}(100)$ keV for the LZ event, we generate this splitting dynamically from the vacuum expectation value (VEV) of a scalar triplet of hypercharge $1$. The same scalar triplet also couples to the lepton doublets while generating light neutrino masses via type-II seesaw mechanism \cite{Mohapatra:1980yp, Schechter:1981cv, Wetterich:1981bx, Lazarides:1980nt}. This not only correlates the small mass splitting in dark sector with the origin of neutrino mass, but also helps reconciling the required inelastic scattering at LZ with the bounds on total DM relic abundance and elastic scattering rate off nucleons. We calculate the event rate at LZ due to up-scattering of SDDM and find the parameter space consistent with the data. While a large part of the model parameter space remains consistent with the LZ data alone, the constraints on elastic DM scattering rate restricts the parameter space strongly.

This paper is organized as follows. In section \ref{sec:model}, we briefly discuss our inelastic SDDM model. In section \ref{sec:LZevent} we discuss the details of the LZ event and its explanation within our SDDM framework. In section \ref{sec:numass}, we briefly comment on the connection to the origin of light neutrino masses and finally conclude in section \ref{sec:conclude}.
\section{Inelastic Singlet-Doublet Dark Matter}\label{sec:model}
We extend the Standard Model (SM) by introducing a vector-like fermion doublet $\Psi=(\psi^0,\,\psi^-)^T$ and a vector-like fermion singlet $\chi$, both odd under a stabilizing $Z_2$ symmetry. The relevant Lagrangian is given as
\begin{align}\label{eq:LagSD}
\mathcal{L}&\supseteq \bar \Psi\big[i\gamma^\mu(\partial_\mu - ig\tfrac{\sigma^a}{2}W_\mu^a
-ig'\tfrac{Y}{2}B_\mu)-m_\Psi\big]\Psi\nonumber\\
		& \quad  +\bar\chi(i\slashed\partial-m_{\chi})\chi -\big(Y\bar \Psi\widetilde H\chi+{\rm h.c.}\big).
\end{align}
After electroweak symmetry breaking, the neutral component of the Higgs field acquires a non-zero VEV: $H\to(0,\,v_h+h)^T/\sqrt2$, and the Yukawa term generates a Dirac-type mass mixing $M_D\equiv Yv_h/\sqrt2$ between $\chi$ and $\psi^0$. Thus, the neutral fermion mass matrix takes the form
\begin{equation}\label{eq:massmatrix}
-\mathcal{L}_{\rm mass}=\overline{(\chi\ \psi^0)}
\begin{pmatrix}m_{\chi} & M_D\\ M_D & m_\Psi\end{pmatrix}
	\begin{pmatrix}\chi\\ \psi^0\end{pmatrix},
\end{equation}
in the flavor basis $(\chi,\,\psi^0)$.
Diagonalizing the above mass matrix we get two mass eigenvalues $m_{\psi_1}$ and $m_{\psi_2}$ given by
\begin{eqnarray}
    m_{\psi_1}\simeq m_{\chi}-\frac{M_D^2}{m_\Psi-m_{\chi}},&\quad
m_{\psi_2}\simeq m_\Psi+\frac{M_D^2}{m_\Psi-m_{\chi}}
\end{eqnarray}
where we have assumed $M_{D}\ll m_{\Psi},m_{\chi}$.
The corresponding mass eigenstates are given by:
\begin{eqnarray}
    \psi_{1}=\cos\theta \chi+\sin\theta \psi^0,\quad
    \psi_{2}=\cos\theta \psi^{0}-\sin\theta \chi,
\end{eqnarray}
where the singlet-doublet mixing angle is given by

\begin{eqnarray}\label{eq:coupling}
    \tan2\theta&=\frac{\sqrt2\,Yv_h}{m_\Psi-m_{\chi}}.
\end{eqnarray} 
Due to the unbroken $Z_2$ symmetry, the lightest $Z_2$-odd particle remains stable. We choose $\psi_1$ to be the lightest, and hence becomes the viable candidate for the dark matter. The mass splitting $\Delta{M}$ between $\psi_1$ and $\psi_2$ is given by $\Delta{M}=m_{\psi_2}-m_{\psi_1}$. From Eq.~\eqref{eq:coupling}, we see that, $Y$ and $\sin\theta$ are not two independent parameters and are related by:
\begin{eqnarray}
    \label{eq:coupsin}Y=\tfrac{\Delta{M}\sin2\theta}{\sqrt{2}v_h}.
\end{eqnarray}
The mass of the charged fermion $\psi^\pm$ in terms of $m_{\psi_1}$, $m_{\psi_2}$ is given by:
\begin{eqnarray}
    m_{\psi^{\pm}}=m_{\psi_1}\sin^2\theta+m_{\psi_2}\cos^2\theta\simeq M_{\Psi}.
\end{eqnarray}
We then introduce a small Majorana mass term $m_1$ ($\mathcal{O}(100)~\text{keV}$) for the neutral component of the fermion doublet $\psi^0$ (see Section~\ref{sec:numass} for details related to the origin of this Majorana mass term from scalar triplet). Since $\psi_1$ is an admixture of the original doublet $\psi^0$ and singlet $\chi$, the $\psi_1$ state splits into two nearly degenerate pseudo-Dirac states $\chi_1,\chi_2$, with masses
\begin{equation}\label{eq:chimasses}
m_{\chi_1}=m_{\psi_1}-m_1,\quad m_{\chi_2}=m_{\psi_1}+m_1.
\end{equation}
The masses of two pseudo-Dirac states are separated by a mass splitting $\delta=2m_1$. The resulting Yukawa, charged and neutral-current interactions of the physical mass eigenstates $\{\chi_1,\chi_2,\psi_2,\psi^-\}$ are
\begin{align}
\label{eq:YukawaChiBasis}
\mathcal{L}_{\text{Yukawa}}
&\supseteq -Y h\bigg[\frac{\sin\theta\cos\theta}{2}\big(\bar\chi_1\chi_1+\bar\chi_2\chi_2\big)\nonumber\\
&\quad-\sin\theta\cos\theta\,\bar \psi_2 \psi_2+\frac{\cos^2\theta}{\sqrt2}\big(\bar \psi_2\chi_1+i\bar \psi_2\chi_2\big)\nonumber\\
&\quad-\frac{\sin^2\theta}{\sqrt2}\big(\bar\chi_1 \psi_2-i\bar\chi_2 \psi_2\big)\bigg]+{\rm h.c.}
\end{align}
\begin{align}
\label{eq:LCC}
\mathcal{L}_{\rm CC}=\frac{e_0}{\sqrt2\sin\theta_W} \left[\frac{\sin\theta}{\sqrt2}\big(\bar\chi_2-i\bar\chi_1\big)\gamma^\mu \psi^-
\right.&\nonumber\\
\left.+\cos\theta\,\bar \psi_2\gamma^\mu \psi^-\right]W_\mu^+ + {\rm h.c.}&
\end{align}
\begin{align}\label{eq:LNC}
\mathcal{L}_{\rm NC} &= g_Z Z_\mu\left\{-i\sin^2\theta\,\bar\chi_1\gamma^\mu\chi_2 +\cos^2\theta\,\bar \psi_2\gamma^\mu \psi_2\right.\nonumber\\
&\quad+\frac{\sin\theta\cos\theta}{\sqrt2}
\Big[\big(\bar\chi_2\gamma^\mu \psi_2+\bar \psi_2\gamma^\mu\chi_2\big) \nonumber\\
&\quad +i\big(\bar \psi_2\gamma^\mu\chi_1-\bar\chi_1\gamma^\mu \psi_2\big)\Big]\left.-\tfrac{\cos2\theta_W}{2}\bar \psi^-\gamma^\mu \psi^-\right\}\nonumber\\
&\quad - e_0\,Q_{\psi^-}\,\bar \psi^-\gamma^\mu \psi^- A_\mu,
\end{align}
where
\begin{equation}\label{eq:gZ}
g_Z\equiv\frac{g}{2\cos\theta_W}=\frac{e_0}{2\sin\theta_W\cos\theta_W}.
\end{equation}
Notably, the neutral-current Lagrangian in Eq.~\eqref{eq:LNC} contains no diagonal $\chi_1$--$\chi_1$ coupling to the $Z$ boson: the Majorana nature of $\chi_1$ forbids such a term identically, leaving only the inelastic $\chi_1\to\chi_2$ transition via Z-exchanged (suppressed by $\sin^2\theta$). From Eq.~\eqref{eq:YukawaChiBasis}, we can see that the DM-nucleon elastic scattering is possible via SM Higgs mediation. However, as we argue in \ref{sec:LZevent} that the elastic Higgs-mediated DM-nucleon scattering is suppressed in comparison to Z-mediated inelastic scattering. In our setup, the relevant parameters for the Direct Detection are $m_{\chi_1}$, $\Delta M$, $\sin\theta$, and $\delta$.

Now we comment on the relevant parameters for estimating DM relic density in our setup. For all practical purposes, we can set $m_{\chi_1}\simeq m_{\chi_2} \simeq m_{\text{DM}}$. Therefore, the relevant parameters for the relic density estimation are $m_{\text{DM}}$, $\Delta{M}$ and $\sin\theta$. By considering annihilation, coannihilation and conversion-driven processes among the states $\psi_1$, $\psi_2$, and $\psi^-$ we can achieve correct relic density in the range $m_{\text{DM}}\sim(100,1000)$ GeV, $\sin\theta\lesssim0.04$ and $\Delta{M}>1$ GeV \cite{Paul:2024prs,Paul:2025spm}.

\section{Inelastic SDDM and recent LZ event}
\label{sec:LZevent}

\begin{figure}[h]
    \centering
    \includegraphics[scale=0.4]{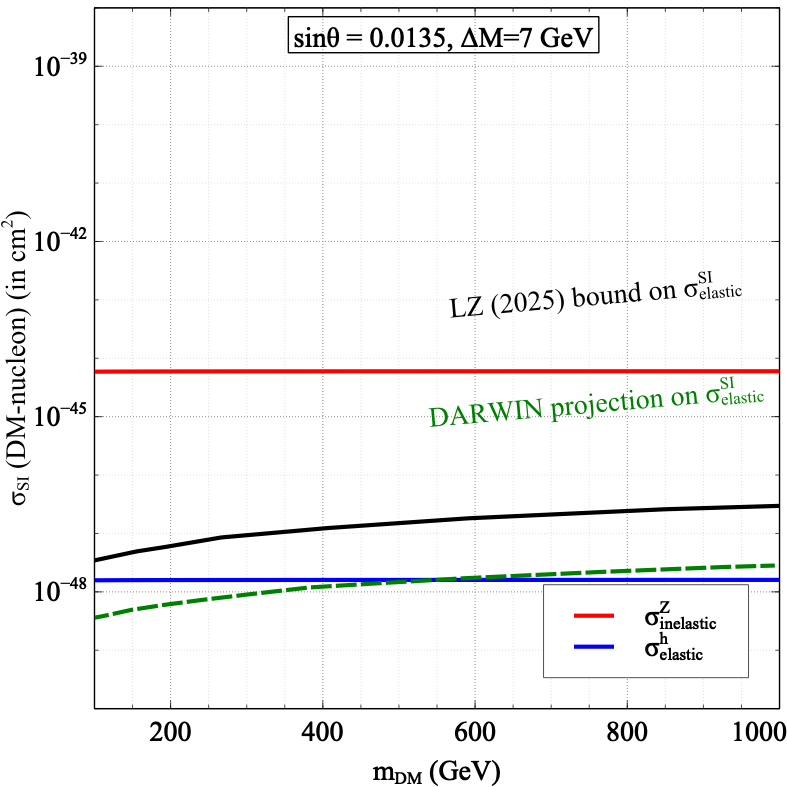}
    \caption{Figure illustrating the suppression of elastic Higgs mediated DM-nucleon scattering compared to inelastic Z-mediated DM-nucleon scattering for a typical benchmark choice: $\sin\theta=0.0135$ and $\Delta M=7$ GeV. The black solid and green dashed contours correspond to LZ 2025 bound \cite{LZ:2024zvo} and projected DARWIN \cite{DARWIN:2016hyl} sensitivity respectively.}
    \label{fig:sigmacomp}
\end{figure}

As discussed in the previous section (see Eq.~\eqref{eq:LNC}), in our setup the scattering of SDDM with nuclei can occur at tree level inelastically through $Z$-boson and elastically via Higgs exchange processes. The relative importance of the two processes can be seen from the corresponding DM-nucleon scattering cross-sections \cite{Djouadi:2011aa},
\begin{eqnarray}
\sigma_n^{Z}&=&\frac{G_F^2\sin^4\theta}{2\pi}\mu_n^2\label{eq:DM_nucleonZ}\\
\sigma_n^{h}&=&\frac{\Delta{M}^2\sin^42\theta}{8\pi v_h^2 M_h^4}\mu_n^2 m_n^2 f_N^2 \label{eq:DM_nucleonH}
\end{eqnarray}
where $\sin\theta$ denotes the singlet-doublet mixing angle, $\Delta{M}$ is the mass splitting between the Singlet-Doublet dark matter states, $\mu_n$ is the DM-nucleon reduced mass, $M_h$ is the Higgs mass, $v_h$ represents the Higgs VEV, $m_n$ is the nucleon mass, and $f_N$ characterizes the Higgs-nucleon coupling. As can be seen from Fig.~\ref{fig:sigmacomp} the contribution from $Z$-boson exchange is dominant compared to Higgs exchange shown for $\Delta{M}=7~\text{GeV},~\sin\theta=0.0135$. At this juncture, we note that the null detection of DM via elastic scattering sets an upper bound on $\sin\theta$ in our model, to be $\sin\theta\lesssim0.04$ for all $\Delta{M}$ satisfying correct relic density \cite{Paul:2024prs}. Therefore, in the rest of the analysis, we will consider only the inelastic Z-mediated scattering for probing the DM at LZ while maintaining $\sin\theta\lesssim0.04$.

\subsection{248 keV LZ event via Z-mediation}
We now consider the kinematic requirements for the LZ event under the assumption that the observed event originates from dark matter-nucleus scattering: $\chi_1 N \to \chi_2 N$ via Z-mediation. For a nuclear recoil of energy $E_R$, the minimum velocity required by an incoming dark matter particle is
\begin{equation}
\label{eq:vmin}
v_{min}=\sqrt{\frac{m_NE_R}{2\mu_N^2}}+\frac{\delta}{\sqrt{2E_Rm_N}},
\end{equation}
where $m_N$ is the nuclear mass, with $m_N \approx 122$ GeV for $^{131}$Xe in the LZ detector, and $\mu_N$ is the reduced mass of the DM-nucleus system. In the elastic limit, $\delta\rightarrow0$, and the minimum velocity increases monotonically with the recoil energy. In contrast, for endothermic scattering, the production of the heavier state $\chi_2$ requires an additional amount of energy. This gives rise to the second term in $v_{min}$, which scales as $\delta/\sqrt{E_R}$ and becomes important for sufficiently small recoil energies.

For a fixed value of $\delta$, the minimum of $v_{min}$ is obtained at
\begin{equation}
\label{eq:ERforvmin}
E_R=\frac{\mu_N}{m_N}\delta.
\end{equation}
Thus, the recoil energy preferred by the inelastic kinematics is directly related to the mass splitting $\delta$, with only a weak dependence on the dark matter mass and the target nucleus through the reduced mass. This relation is useful for understanding why a relatively large recoil energy can be associated with a particular range of mass splittings.

To describe the dark matter velocity distribution, we adopt the Standard Halo Model (SHM), in which the dark matter velocities in the Galactic frame follow a truncated Maxwell-Boltzmann distribution \cite{Lewin:1995rx},
\begin{equation}
\label{eq:velocitydistn}
f_{\text{SHM}}^{\text{gal}}=k e^{-v^2/v_0^2}\Theta(v_{\rm esc}-v),
\end{equation}
where $k^{-1}=(\pi v_0^2)^{3/2}\left[\text{erf}(\frac{v_{esc}}{v_0})-\frac{2}{\sqrt{\pi}}\frac{v_{esc}}{v_0}e^{-v_{esc}^2/v_0^2}\right]$, $v_0 = 220$ km/s is the velocity dispersion, $v_{\rm esc} = 540$ km/s is the Galactic escape velocity, and $\Theta$ denotes the Heaviside step function.

The differential nuclear recoil rate at a direct detection experiment can then be written as
\begin{equation}
\label{eq:dRdER}
\frac{dR}{dE_R}=\frac{\rho_{\text{DM}}}{m_\text{DM}}N_T\int_{v>v_{min}}v f^{\text{gal}}(\vec{v}+\vec{v_e}(t))\frac{d\sigma}{dE_R}d^3v,
\end{equation}
where $\vec{v}$ is the dark matter velocity in the Earth frame, $\vec{v}_e(t)$ is the velocity of the Earth with respect to the Galactic frame, $\rho_{\text{DM}} \approx 0.4$ GeV/cm$^3$ is the local dark matter energy density. We assume that SDDM constitutes the entire dark matter abundance. Here, $N_T$ is the number of scattering targets per unit detector mass. The spin-independent differential cross section for scattering off a nucleus is given by
\begin{equation}
\label{eq:dsigmadER}
\frac{d\sigma}{dE_R}=\frac{\sigma_{n}}{v^2}\frac{m_N}{2\mu_n^2}\left(\frac{Zf_p+(A-Z)f_n}{f_n}\right)^2 F^2(E_R),
\end{equation}
where $\sigma_n$ and $\mu_n$ are respectively the DM-nucleon scattering cross section, given in Eq.~\eqref{eq:DM_nucleonZ}, and the corresponding reduced mass. The quantities $A$ and $Z$ denote the nuclear mass and atomic numbers, respectively, while $f_p$ and $f_n$ are the effective DM-proton and DM-neutron couplings. The nuclear structure effects are included through the form factor $F(E_R)$, which we take to be the Helm form factor \cite{Engel:1991wq,Duda:2006uk},
\begin{equation}
\label{eq:Helmformfactor}
F^2(E_R)=\left(\frac{3j_1(qr_0)}{qr_0}e^{\frac{-(qs)^2}{2}}\right)^2.
\end{equation}
where, $j_1(x)$ is the spherical Bessel function of the first kind, $q=\sqrt{2m_N E_R}$, $s\simeq0.9$ fm, $r_0=\sqrt{c^2+\frac{7}{3}\pi^2a^2-5s^2}$, with $a\simeq0.52$ fm, and $c=1.23 A^{1/3}-0.6$. Using the above ingredients, we perform a simple test of the SDDM parameter space that can potentially account for the LZ event. For this purpose, we employ the extended maximum likelihood method \cite{BARLOW1990496}. The extended likelihood is defined as a function of the model parameters ${x_i}$:

\begin{eqnarray}
{\mathcal L}({x_i}) & \equiv & \left[\prod_{i=1} P(x_i)\right]e^{-\mathcal{N}}, \nonumber \\
& = & \left[\prod_{i=1}^{n_o} \left.\frac{d N({x_i})}{d E_R^\prime}\right|_{E_R^\prime=E_i}\right]e^{-\mathcal{N}({x_i})},
\end{eqnarray}
where $\mathcal{N}= \int_{E_R^{\rm min}}^{E_R^{\rm max}} \frac{d N({p})}{d E_R^\prime} d E_R^\prime~,$ represents the total number of events predicted over the recoil-energy interval $\left[E_R^{\rm min},E_R^{\rm max}\right]$ for a particular choice of model parameters. The differential number of events is obtained from the differential recoil rate as $dN/dE_R=dR/dE_R,\times$ experimental exposure, where the LZ exposure is 2.84 tonne-year. The quantity $n_o$ denotes the number of observed signal events, with each event occurring at a recoil energy $E_i$. In the present analysis, the relevant model parameters are $m_\text{DM}$, $\delta$, and the mixing angle $\sin\theta$. We consider a single observed signal event, i.e., $n_o=1$, with recoil energy
$E_i=248\pm 23$(stat)$\pm 23$(sys) keV. We take the recoil-energy range probed by LZ to be $E_R^{\rm min}=5.4$ keV to $E_R^{\rm max}=269.9$ keV.

\begin{figure}[h]
\centering\includegraphics[scale=0.4]{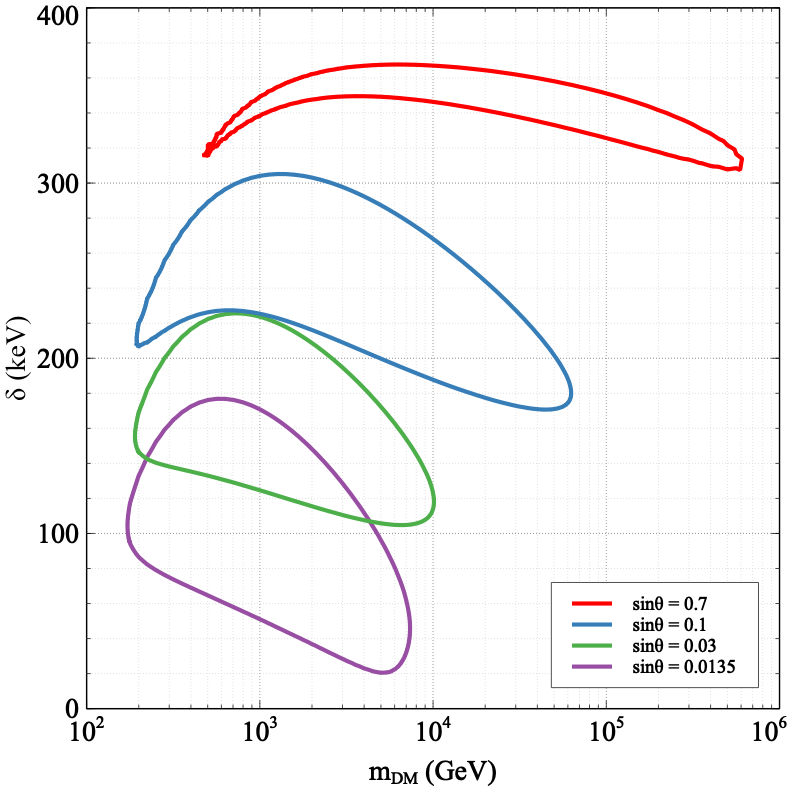}
    \caption{$1\sigma$ contours that could account for the LZ event in the ($m_\text{DM}$,$\delta$) plane with the Singlet-Doublet inelastic Dark Matter with mixing angle $\sin\theta=0.0135,~0.03,~0.1,$ and $0.7$ (purple, green, blue and red contours respectively).}
    \label{fig:contour01}
\end{figure}

We evaluate the likelihood $\mathcal{L}(m_\text{DM},\delta)$ over the parameter space of interest and determine its maximum numerically,
\begin{eqnarray}
\ln\mathcal{L}_{\rm max}
&=& \max_{m_\text{DM},\delta}\ln\mathcal{L}(m_\text{DM},\delta)\\
&=& \max_{m_\text{DM},\delta}
\left(-\mathcal{N}
+\ln\left(\left.\frac{dN}{dE_R}\right|_{E_R=E_i}\right)\right).\nonumber
\label{eq:maxlikelihood}
\end{eqnarray}
The corresponding values $(m_\text{DM}^{\rm bf},\delta^{\rm bf})$ define the best-fit point. We characterize the region preferred by the event using the likelihood-ratio statistic
\begin{equation}
\Delta\chi^2(m_\text{DM},\delta)
=-2\left[\ln\mathcal{L}(m_\text{DM},\delta)
-\ln\mathcal{L}_{\rm max}\right].
\label{eq:deltachi2}
\end{equation}
For two simultaneously varied parameters, we define the $1\sigma$ region using the standard likelihood-ratio criterion
\begin{equation}
\Delta\chi^2(m_\text{DM},\delta)\leq 2.30,
\label{eq:contourdef}
\end{equation}
corresponding to a $68.3\%$ confidence region under the usual $\chi^2$ approximation. The $1\sigma$ contours shown in Fig.~\ref{fig:contour01} are therefore obtained by numerically tracing the curve $\Delta\chi^2=2.30$ in the $(m_\text{DM},\delta)$ plane for different values of $\sin\theta$.

Following Eq.~\eqref{eq:DM_nucleonZ}, we see that as $\sin\theta$ decreases, $\sigma^Z_n$ falls steeply, as the fourth power of $\sin\theta$. To continue reproducing an event rate compatible with the single observed LZ event, the fit compensates by favoring smaller $\delta$ and smaller $m_\text{DM}$: a smaller $\delta$ reduces $v_{\rm min}$, while a smaller mass directly enhances the number density $\rho_{\text{DM}}/m_\text{DM}$. This is precisely the trend visible in Fig.~\ref{fig:contour01}, i.e., reducing $\sin\theta$ from $0.7$ to $0.0135$ shifts the preferred mass splitting from $\delta \sim 310$--$370\,\mathrm{keV}$ down to $\delta\sim30$--$180\,\mathrm{keV}$, and the preferred mass range down from $m_{\rm DM}\sim10^{4}$--$10^{5.8}\,\mathrm{GeV}$ to $m_{\rm DM}\sim10^{2.5}$--$10^{3.8}\,\mathrm{GeV}$. The contours also shrink in extent as $\sin\theta$ decreases, reflecting the narrowing range of $(m_{\text{DM}},\delta)$ combinations capable of compensating the falling cross section.

At this point, it is worth shedding light on our parameter choice based on other phenomenological bounds on the model parameter space. Earlier analyses in Refs.~\cite{Bhattacharya:2018fus,Paul:2025spm} show that the DM relic density can be satisfied over a wide range of DM mass, from $\mathcal{O}(10^2)$ GeV to $\mathcal{O}(10^3)$ GeV, with $\sin\theta$ ranging from $\mathcal{O}(10^{-7})$ up to maximal allowed mixing, i.e., 0.04 (relic density + Direct detection). The DM relic density is also found to have negligible dependence on the mass splitting $\delta$ in this model. We show that correct relic density can be satisfied for the benchmark points $\{m_{\text{DM}},\Delta M,\sin\theta\}$ considered in the Fig.~\ref{fig:contour01} which are given as:
\begin{eqnarray}
    \text{BP1: }&& \{800~\text{GeV},2.5~\text{GeV},0.7 \}\nonumber\\
    \text{BP2: }&& \{700~\text{GeV},1.5~\text{GeV},0.1\},\nonumber\\
    \text{BP3: }&& \{700~\text{GeV},1.2~\text{GeV},0.03\},\nonumber\\
    \text{BP4: }&& \{700~\text{GeV},1.2~\text{GeV},0.0135\}\nonumber
\end{eqnarray}
Using the Eq.~\eqref{eq:DM_nucleonH}, we find the direct detection DM-nucleon elastic cross-section for each benchmark point as $(\text{BP1:}~5.38\times10^{-43} \text{cm}^2,~\text{BP2:}~3.04\times10^{-46}\text{cm}^2,~\text{BP3:}~1.60\times10^{-48}\text{cm}^2,~\text{BP4:}~6.58\times10^{-50}\text{cm}^2~)$. Thus we find that BP3 and BP4 are safe from the elastic spin-independent LZ bound \cite{LZ:2
024zvo}.

\section{Inelastic SDDM and connection to Neutrino Mass}\label{sec:numass}
In Section~\ref{sec:model}, we briefly introduced a Majorana mass for the doublet component $\psi^0$ as a prerequisite for generating the pseudo-Dirac fermion states, and later showed that this is precisely what forbids $Z$-mediated elastic DM scattering ($\chi_1N\to\chi_1N$) while opening up an inelastic transition to a heavier state ($\chi_1N\to\chi_2N$). Such a Majorana mass term, however, is not gauge invariant under the electroweak symmetry. In this section, we present a mechanism that generates it dynamically: we extend the model with a scalar triplet field $\Delta\sim(\mathbf{1},\mathbf{3},1)$, taken to be trivial (even)
under $Z_2$. The scalar triplet is represented in matrix notation as:
\begin{equation}
\Delta = \begin{pmatrix} \Delta^+/\sqrt{2} & \Delta^{++} \\ \Delta^0 & -\Delta^+/\sqrt{2} \end{pmatrix}.
\end{equation}
The extended Lagrangian can be written as
\begin{align}
\label{eq:Lagtriplet}
\mathcal{L}&\supseteq  -\frac{1}{\sqrt{2}}y_\Delta\Big[\overline{\Psi^c}(i\sigma_2\Delta)\Psi+{\rm h.c.}\Big]\nonumber\\
&\quad-\frac{1}{\sqrt2}(y_L)_{\alpha\beta}\Big[\overline{L_\alpha^c}(i\sigma_2\Delta)L_\beta+{\rm h.c.}\Big]-V(H,\Delta),
\end{align}
where the first term is the triplet-doublet Yukawa coupling responsible for generating the Majorana mass $m_1$ on $\psi^0$ discussed in Section~\ref{sec:model}, while the second term is the familiar Type-II seesaw coupling of $\Delta$ to the SM lepton doublets $L_\alpha=(\nu_\alpha,\,e_\alpha)^T$, responsible for generating Majorana neutrino mass. $V(H,\Delta)$ is the scalar potential for the SM Higgs doublet $H$ and the scalar triplet $\Delta$.
\begin{align}
V(H,\Delta) &= -\mu_H^2 H^\dagger H+\lambda_H(H^\dagger H)^2+M_\Delta^2\,{\rm Tr}(\Delta^\dagger\Delta)\nonumber\\
&\quad+\lambda_\Delta\big[{\rm Tr}(\Delta^\dagger\Delta)\big]^2+\lambda_{\Delta2}\,{\rm Tr}\big[(\Delta^\dagger\Delta)^2\big]\nonumber\\
&\quad+\lambda_3(H^\dagger H)\,{\rm Tr}(\Delta^\dagger\Delta)+\lambda_4\,H^\dagger\Delta\Delta^\dagger H\nonumber\\
&\quad+\big(\mu\,H^\dagger\Delta\widetilde H+{\rm h.c.}\big),
\end{align}
with $\widetilde H=i\sigma_2H^*$.
After electroweak symmetry breaking, the scalar triplet acquires an induced VEV:
\begin{equation}
v_\Delta \approx -\frac{\mu v_h^2}{2 M_\Delta^2 + (\lambda_{3} + \lambda_{4}) v_h^2}
\end{equation}

This gives not only Majorana mass to the SM neutrinos: $(M_\nu)_{\alpha\beta}=\sqrt{2}(y_L)_{\alpha\beta}~v_\Delta$, but also a Majorana mass $m=y_N ~v_\Delta$ to the neutral state $\psi^0$ of the doublet fermion. As a result, through the singlet-doublet mixing, $\psi_1$ gets a Majorana mass $m_1=m~\sin^2\theta$.

\section{Conclusion}
\label{sec:conclude}
We have revisited one of the popular WIMP DM scenarios namely, the singlet-doublet fermion DM model to check the possibility of explaining the recently reported high-energy nuclear recoil event by the LZ collaboration. The Dirac fermion mass eigenstates of SDDM are split into pseudo-Dirac components by a scalar triplet which is also responsible for generating light neutrino masses by the type-II seesaw mechanism. This allows the up-scattering of the DM into the next-to-lightest component via Z-mediated process, which can give rise to the nuclear recoil event reported by LZ. We show that the model can explain the LZ event for a wide range of DM mass, pseudo-Dirac mass splitting as well as singlet-doublet mixing angle. However, the constraints from elastic DM-nucleon scattering forces the parameter space to a narrower parameter space in terms of DM mass, pseudo-Dirac mass splitting and singlet-doublet mixing angle. As the singlet-doublet mixing angle controls both the inelastic and elastic scattering rates, future constraints on the latter can further narrow down the parameter space of our model. Apart from the usual phenomenology of SDDM, the presence of scalar triplet provides a connection to the origin of neutrino masses while offering additional phenomenology as well as detection avenues.

\section*{Acknowledgments}
The work of D.B. is supported by the Science and Engineering Research Board (SERB), Government of India grant CRG/2022/000603. S.K.S acknowledges the support provided by IMSc, Chennai, during his visit.

\appendix

%

\end{document}